\documentstyle[twocolumn,aps]{revtex}

\begin{document}
\twocolumn[
\hsize\textwidth\columnwidth\hsize\csname@twocolumnfalse\endcsname
\draft
\title{\bf Hidden spin-current
conservation in {\bf $2d$} {\bf Fermi liquids}}
\author{Paulo F. Farinas}
\address{Instituto de F\' \i sica Gleb Wataghin,
Universidade Estadual de Campinas 13083-970,
Campinas, S\~ ao Paulo, Brazil}
\author{Kevin S. Bedell}
\address{Department of Physics, Boston College,
Chestnut Hill, Massachusetts 02167}
\author{Nelson Studart}
\address{\it Departamento de F\'{\i }sica, Universidade Federal
de S\~{a}o Carlos 13565-905, S\~{a}o Carlos, S\~{a}o Paulo, Brazil}

\date{\today}
\maketitle

\begin{abstract}
We report the existence of regimes of the two dimensional Fermi liquid that
show unusual conservation of the spin current and
may be tuned by varying some parameter like the density
of fermions.
We show that for reasonable models of the effective interaction the
spin current may be conserved in general in $2d$, not
only for a particular regime. Low temperature spin waves
propagate distinctively
in these regimes and entirely new
``spin-acoustic'' modes are predicted for scattering-dominated temperature
ranges. These new high-temperature propagating spin waves provide a
clear signature for the experimental search of such regimes.
\end{abstract}
]

Fermi-liquid theory (FLT) was first introduced by Landau\cite{Landau}
and further
developed to include spin waves a long time ago.\cite{Silin}
The progress of FLT in $3d$ revealed that it is one of
the broader theories in condensed matter physics, explaining
the experimental results of a wide range of different
systems. Recently,
interest in FLT re-gained momentum,
driven in part by the discovery of high temperature
superconductors and also by the refinement of experimental
techniques in low dimensional physics. While many results on
the (ab)normal phases of the
former allows one interpretation that casts
doubts on the validity of FLT in $2d$,\cite{Anderson}
the latter has been consolidating a source of examples
of practical $2d$ systems that behave as predicted
by $2d$ FLT, as it can be appreciated in the
experiments reported in $^{3}$He
films.\cite{sixPcent,Hallock}
This seems also to be the case for 
doped semiconductors, where thickness and doping
can be controlled. Experiments on these charged
systems have directly observed
expected $2d$ Fermi liquid behavior in GaAs heterostructures.%
\cite{Murphy} This comes from extracting the quasi-particle
life times from the tunneling peaks in the current-voltage profile
of two biased $2d$ doped semiconductor contacts with
a quantum well between them. The
result is the one predicted by $2d$
FLT.\cite{Hodges}

Spin waves were observed in bulk alkali metals
some three decades ago by conduction-electron spin
resonance (CESR) techniques,\cite{Schultz} confirming
the predictions of FLT.\cite{Platzman}
Later, using nuclear magnetic resonance (NMR) in bulk $^3$He,
Ref.\cite{Corruccini}
confirmed the existence of the Leggett-Rice effect
predicted formerly.\cite{Leggett} The suppression of the Leggett-Rice
effect was also indirectly observed for a particular
region of the parameter space in bulk $^{3}$He-$^{4}$He
mixtures,\cite{Bradley,Masuhara,Ishimoto}
which was pointed out in Ref.\cite{BedellPRL}
All these results agree with the known
fact that spin-current is not conserved in $3d$.\cite{BaymPethick}
The majority of these experiments
were not repeated for $2d$ systems up to this date.

The central purpose of this Letter is to present exact results
indicating that spin-current may be a conserved
quantity in $2d$ Fermi liquids, at least for some regions
of the parameter space. 
We will see that for microscopic models that assume
short range potentials for the effective
interaction, spin-current
conservation holds for the entire parameter space, so that
such models validate the bold statement that spin current is conserved
in $2d$ (not only at particular regimes). This is rather compelling
since our results are for the spin channel, and the transition
from short to long range in the charge channel should not essentially modify
the physics presented here, provided one has no broken symmetries.
We also calculate the dispersions of transverse spin-waves under
this ``spin-space Galilean
invariance'' and find that experimental observation of
these new collective modes
will show rather distinct features that 
make such regimes easy to identify. Before proceeding, we should
like to give a precise meaning for $2d$ in the
context of this article. Let us estabilish that
the system is $2d$ whenever one of its $3$ dimensions is
comparable to or less than the quasiparticle's typical
wavelength.

At low temperatures, the phase space available for scattering in $3d$ is a
spherical shell and the incoming momenta are not in general co-planar with
the outgoing pair. As a result, the value of the spin current carried
by two quasiparticles with anti-parallel spins (singlet channel),
before and after they collide are not related, ${\bf j}%
_{in}\equiv \sigma _{1}{\bf p}_{1}+\sigma _{2}{\bf p}_{2}=%
{\bf p}_{1}-{\bf p}_{2}%
\neq {\bf j}_{out}={\bf p}_{3}-{\bf p}_{4}$,
where $1,2$ and $3,4$ refer to
incoming and outgoing momenta. If ${\bf q}$ is the total exchange of
momentum in the collision, we can write
$%
{\bf j}_{out}={\bf j}_{in}+{\bf q}(\phi )$,
where $\phi $ is the angle between the scattering planes. While the triplet
channel conserves spin current trivially, the two scattering processes in
the singlet
channel are completely accounted for by fixing the spins on all momenta
and varying $\phi $. One can turn from small momentum exchange, near $\phi =0$
(forward), continuously into large $q \sim 2k_{F}$, near $\phi =\pi $
(backward). Since ${\bf j}_{out}$ points in a random
direction relative to ${\bf j}_{in}$, we can say that the
spins ``walk'' randomly throughout the system and hence spin
transport is diffusive. In $2d$, however, due to the reduced phase space,
one is left with only three possibilities, ${\bf j}_{out}={\bf j}_{in},$ $%
{\bf j}_{out}=-{\bf j}_{in},$ and ${\bf j}_{in}({\bf k}=0)\neq {\bf j}_{out}(%
{\bf k}=0),$ where ${\bf k}$ is the total momentum. This latter region of
the phase space brings no contributions to the scattering integral to
leading order in temperature.\cite{Hodges,Ebner,Mula} In fact, as we
see below, the scattering amplitude for ${\bf k}=0$ processes is zero in
a regime that conserves spin-current, so that
we are left with only two scattering processes:
forward that conserves spin current and backward that flips the
direction of spin current. Hence, spin-current
is not conserved in $2d$ as long as the balance between
two clearly distinguishable
processes remains.

To study the circumstances that allow this balance to
break in such a way as to favor spin-current conservation,
we consider a planar Fermi liquid with a weak
magnetization perpendicular to the plane and whose gradient is
such that ${\bf \nabla}M = |{\bf \nabla}M|\hat{z}$, where
$\hat{z}$ is an in-plane unit vector.
The scattering integral for the Landau kinetic equation of FLT
in $2d$ can in general be expanded in circular harmonics, $\psi_l(x)$ ,
\[
I[n_{{\bf p}\sigma }] = %
\sum_{l}I_{l}\psi_{l}({\bf p}\cdot \hat{z})\;,
\]
where the amplitudes depend on the quasiparticle energy. 
For spin-diffusion processes,
while $I_{0}=0$ due to spin conservation in collisions,
the symmetry of the distribution function implies that all but the first
higher order
angular contributions are small enough so that we can write
$I_{l}=I_{1}\delta_{l1}$.
In $2d$ this dominant amplitude can be written explicitly
in terms of the low-frequency four point vertex function
$\Gamma $ at $k=0$
and $k=2k_{F}$,
\begin{eqnarray}
I_{1} = C\left[u_{1}(x_{\bf p})S + u_{2}(x_{\bf p})%
\left|\Gamma_{\uparrow%
\downarrow \downarrow \uparrow}^{0}\right|^{2}\right]\;, \label{sctint}
\end{eqnarray}
where
\[
C\propto \sigma |{\bf \nabla}%
M| \left(\frac{T}{T_F}\right)^{2}%
 \left|ln\left(\frac{T}%
{T_F}\right)\right|\;,
\]
\[
S=\sum_{spins,k=0,2k_{F}}%
\left|\Gamma_{\sigma_{1}%
\sigma_{2} \sigma_{3} \sigma_{4}}^{k}\right|^{2}%
\delta_{\sigma_{1}+\sigma_{2},\sigma_{3}+\sigma_{4}}\;,
\]
and $u_{i}$ are functions of the normalized energy
$x_{\bf p}={(\epsilon_{\bf p}-\mu)}/k_{B}T$. Equation
(\ref{sctint}) yields 
\begin {equation}
\sum_{\sigma {\bf p}}\sigma {\bf p}I[n_{{\bf p}\sigma}]%
 \propto \left|\Gamma_{\uparrow%
\downarrow \downarrow \uparrow}^{0}\right|^{2}%
\;. \label {conserv}
\end {equation}
This result comes
from the vanishing of the term proportional to $S$
in Eq.(\ref{sctint}) under energy integration.
Details will be presented elsewhere but they follow from a similar
analysis of the scattering integral in $2d$ found in Refs.\cite{Hodges}
and \cite{Mula}.
The fact that in $2d$ one can write exact expressions in terms
of the Landau parameters for
the vertex part \cite{dial}
yields closed expressions in terms of this
function at particular points of the phase space. 
It is clear from Eq.(\ref{conserv}) that spin current is
conserved {\it exactly} if
$\Gamma_{\uparrow \downarrow \downarrow \uparrow}^{0}=0$.
In the same spirit,
the diffusion coefficient may also be expressed
in terms of the vertex part\cite{Mula} for a weakly polarized
Fermi liquid, and when
$\Gamma_{\uparrow \downarrow \downarrow \uparrow}^{0}=0$
it diverges, as it should be if spin current is
conserved.

To see that
$\Gamma_{\uparrow \downarrow \downarrow \uparrow}^{0}$ may
in fact be zero,
we start with a dilute-gas result\cite{Jan} that
gives
$\Gamma_{t-matrix}^{k=0}=0$. This result 
comes from re-summing the logarithm divergences
for an arbitrary spin independent short range potential,
similarly to what is done in $3d$.\cite{Galitski}
This is a physically compelling
starting point, since, as mentioned earlier,
one does not expect the physics
in the spin channel to radically change in the absence
of broken symmetries for potentials with longer tails
in the charge channel.
We plug the t-matrix
result for an arbitrary ${\bf k}$ in the Bethe-Salpeter
equation\cite{dial} for the vertex function
and obtain
$\Gamma_{\uparrow \downarrow \downarrow \uparrow}^{0}=0$.
This is equivalent to including contributions
from RPA diagrams to all orders in the
coupling constant ($g$)
in addition to particle-particle
ladder bubbles to all orders and particle-hole
ladder bubbles up to second order that are included in
$\Gamma_{t-matrix}^{k}$.
Contributions from other diagrams,
if not vanishing at the particular point $k=0$, will be
small, leading to a very long diffusion relaxation time.

For this result
the spin-diffusion relaxation time due to $I_{1}$ is infinite
so that the symmetry due to higher
order amplitudes becomes dominant. 
We stress that even
in a more general scenario where 
$\Gamma_{\uparrow \downarrow \downarrow \uparrow}^{0}$
remains small but finite, it suffices that it is small enough
to render the relaxation time associated with $I_{1}$
long compared to the scales arising
from higher angular terms. Under such condition 
spin-current will be conserved
in the relevant finite time scales, the system will
relax due to a higher order process that we call ``spin-viscosity''
in analogy to what happens with sound.

We discussed the possibility that a new conservation
exists in $2d$ Fermi liquids in general. This is
not the only possibility. We now
turn to a more phenomenological analysis of
particular regimes that conserve spin-current in $2d$.
This is based on the fact that the condition for
$\Gamma_{\uparrow \downarrow \downarrow \uparrow}^{0}=0$, if
not valid in general in $2d$, can be achieved {\it by tuning}
appropriate values of the Landau parameters. To be more
specific, let $a$ be the Greens' function quasiparticle residue
and $N(0)$ the density of states at the Fermi surface.
For a $2d$ Fermi liquid we can write\cite{Mula,dial}
\begin{equation}
N^2(0)a^2%
|\Gamma_{\uparrow \downarrow \downarrow \uparrow}^{0}|^2=%
\left[\sum_{l=-\infty}^{+\infty}(-1)^{l}c_{l}A_{l}^{a}%
\right]^2\;, \label{scampl}
\end{equation}
where $A_{l}^{a}$ are the antisymmetric scattering amplitudes,
that relate to the Landau coefficients through
$A_{l}^{a} = F_{l}^{a}/(1 + c_{l}F_{l}^{a})$, and
$c_{l} = 1/(2|l| + 1)$. A quick technical remark about our
choice of the $2d$ basis will avoid confusion. We choose
$\psi_{l}({\bf p}\cdot \hat{z}) = c_{l}e^{il\theta}$.
This shifted basis leads mostly to formulas that look identical
to their $3d$ counterparts, and is only a matter
of convenience. We see from Eq.(\ref{scampl}) that regions of
the parameter space that conserve spin current {\it exactly}
are tuned when $c_{l}A_{l}^{a}=2^{\delta_{l0}}c_{l+1}A_{l+1}^{a}$,
for $l=0,2,4,...$,
$c_{l}A_{l}^{a}=2^{\delta_{l0}}c_{l+3}A_{l+3}^{a}$, for
$l=0,1,2,...$, and an infinite number of other possibilities.
One can also keep only the first few scattering amplitudes following
another dilute-gas calculation for which the $n$-th
Landau coefficient is proportional to $g^n$.\cite{saco}
One then finds that
$\Gamma_{\uparrow \downarrow \downarrow \uparrow}^{0}=0$ for
specific combinations of the Landau parameters.
For instance, when the first two Landau parameters are such that
$F_0^a = 2F_1^a/3(1 - F_1^a/3)$, the contribution from the
first two terms in the sum vanishes and one is left only
with terms that are of order $g^2$ or higher. Throughout this
article, the $2d$ coupling constant,
$g\equiv -1/2ln(k_{F}a_s)$, arises from microscopic models based
on short-range potentials with characteristic
length $a_s$. One can easily verify that, for the majority of
choices made to give
$\Gamma_{\uparrow \downarrow \downarrow \uparrow}^{0}=0$,
one always finds a very reasonable value for $g$ (between
$0.1$ and $0.4$). It is clear that these regimes may be
tuned by externally varying some parameter such as the density.
This particular range of the coupling constant corresponds
to second-layer coverage densities between $5$ and $18\times 10^%
{13}$cm$^{-2}$ in the data from $^3$He films on Grafoil
experiments.\cite{grey}

Given the infinite number of combinations that lead to
$\Gamma_{\uparrow \downarrow \downarrow \uparrow}^{0}=0$,
we feel compelled to investigate the signatures that
should indicate the
experimental tuning of such regimes in the spin-wave modes.
The dispersion relations for transverse spin-waves are the
simpler and broader objects that we can think of for this
purpose. They may be used to figure out the form of the
effective ``diffusion'' (in this case we should say
effective ``viscosity'') equivalent to the Platzman and Wolff
result for charged systems so that CESR experiments can
be done in electronic planes.
We will hence leave the consequences on
more sophisticated phenomena
like the Leggett-Rice effect\cite{Leggett} for future publications.
Also, the detailed analysis of the dispersion
relations will be published in a longer
article, here we only show the results and outline the derivation. 

Spin current conservation equalizes the spin and charge channels
regarding the number of conserved quantities. As a consequence, we expect
propagation of spin-waves to become analogous to sound
propagation. This is partly true, the presence of an
additional scale set by
the external magnetic field keeps the propagation of spin waves distinct
from sound but they present various new common features.
We project out Landau
kinetic equation\cite{BaymPethick}
on the $2d$ basis, and solve for
the Fermi-surface distortions associated with
transverse spin-waves in a relaxation time scheme.
For
$\Gamma_{\uparrow \downarrow \downarrow \uparrow}^{0}=0$,
the relaxation time approximation is written as
$I_{l} = (1-\delta_{l0}-\delta_{l1})\omega_{\sigma \eta}$, where
we have introduced the spin-viscous relaxation time
$\omega_{\sigma \eta}^{-1}$.  
The dispersion relations are calculated for a weak
but finite magnetic
field such that in the long wavelength limit 
$qv_F\ll \omega _L$, the Larmor frequency.
For bulk $^3$He this corresponds
to magnetic fields between $0.25$ and $1$ Tesla. 
The results are, 
\[
\Delta \omega _{l=0}=-\frac{\alpha _0}{2\lambda \omega _L}(qv_F)^2+\left(%
\frac 1{\alpha _1}-\frac{\lambda \lambda _2}{2z}\right) \frac{\alpha _0^2}{%
4\lambda ^3\omega _L^3}(qv_F)^4\;,
\]
\[
\Delta \omega _{l=1}=\omega _L\lambda \alpha _1+\left[ 1+\frac{(1+\lambda%
_2\alpha _2)}{2(1-z\alpha _2/\lambda \alpha _1)}\right] \frac{\alpha _0}{%
2\lambda \omega _L}(qv_F)^2\;,
\]
and 
\[
\Delta \omega _{l=2}=\omega _Lz\alpha _2-\frac{(1+\lambda _2\alpha _2)}{%
2(1-z\alpha _2/\lambda \alpha _1)}\frac{\alpha _0}{2\lambda \omega _L}%
(qv_F)^2\;,
\]
where $\alpha_{l}=1+c_{l}F_{l}^{a}$.
In addition to the usual strength $\lambda \equiv%
\alpha_{0}^{-1}-\alpha _{1}^{-1},$ we
defined $\lambda _2\equiv%
\alpha _0^{-1}-\alpha _2^{-1}$.  
Here $\Delta \omega =\omega -\omega _L$ and $z\equiv%
\lambda _2-i\omega_{\sigma \eta }/\omega _L$.
We write the lowest distortions to higher order in $q$ since
these modes attenuate only to relative order $q^2$ under the
new condition of spin-current conservation.
A third dispersion branch emerges as a result of
the higher order ``spin-viscous'' attenuation.
We see that the quadrupolar fluctuations show similar
behavior to the lower-order distortions in a
non-conserving regime: it
propagates at low $T$ and is purely damped at high $T$.
The dipole modes, however,
propagate almost undamped both at high and low $T$'s, showing
more attenuation for intermediate temperatures. The density fluctuation
propagates almost undamped and with the same dispersion for any $T$
to leading order in $q$, and is weakly damped to relative order $q^2$ at
intermediate temperatures. Hence the $l=0$ and $l=1$ modes propagate both in
the collisionless and hydrodynamic regimes. This rather unusual behavior in
the propagation of the two first spin-wave modes is a direct consequence of
spin current conservation and it is analogous to what happens in
the propagation of sound. The ratio $\omega%
_{\sigma \eta }/\omega _L$ governs the interplay between the collisionless and
hydrodynamic regimes and a
peak in the attenuation occurs when $\omega_{\sigma \eta%
}/\omega _L=\lambda _2$ for both modes.

For completeness, we address some regimes of the parameter space
that are of particular interest if they co-exist with
conservation of spin current. We do not
wish to imply here that these additional regimes exist, only to
present the changes that should be expected if 
they do. First we
look at the regime for which $\lambda = 0$. This regime
where the interaction strength changes sign
was observed in bulk $^3$He-$^4$He mixtures,
\cite{Bradley,Masuhara,Ishimoto} and it is known to show a
remarkable suppression of the Leggett-Rice effect.\cite{Leggett,BedellPRL}
For $\lambda =0$, the relevant changes occur in the lowest two
branches which
collapse into two physically equivalent branches corresponding to sound-like
spin waves propagating in opposite directions with velocity $\alpha _0v_F/ 
\sqrt{2}.$ These modes propagate at any temperature
to leading order in $q$. The attenuation in this case is of
relative order $q$ and also presents a peak at
$\omega_{\sigma \eta }/\omega _L=\lambda_2$. These modes are thus
analogous to sound in the sense that besides having a linear dispersion 
they undergo a transition from a low temperature
zero-sound-like regime into a hydrodynamic regime
as one raises the temperature.

The region of the parameter space for which $\lambda_2=0$ is readily
obtained and brings no additional physics if $%
F_0^a\neq F_1^a/3$ unless for the fact that the attenuation of
the $l=0$ mode becomes of relative order $q^4$. However, for
$\lambda =\lambda_2 = 0$ we have the interesting new feature that the
attenuation of the two lowest order distortions decrease as $T$
increases. This may be understood by recalling that the strength of the
quasiparticles' interactions is measured by $\lambda $ and $\lambda _2$. If
both parameters are zero then no zero-sound-like modes are expected to
propagate at low $T$. However, as scattering increases with $T$,
spin currents are favored in a conserving regime.

In conclusion, we presented the possibility that
spin-current is conserved in $2d$ Fermi liquids,
if not in general, for some particular regimes.
This is basically due to the
restricted geometry combined with the degeneracy of the Fermi surface.
We showed some consequences this conservation brings to the
propagation of transverse spin waves in such regimes and predicted that
spin waves, that are known to occur only for very low $T$,
will propagate also in scattering dominated regimes. This is the most
remarkable property of such regimes and should serve
to clearly distinguish what we call
``spin-viscous'' relaxation processes
associated with such regimes from
ordinary spin-diffusion relaxation. NMR experiments probing
the spin-relaxation will show sharper absorption
peaks for the two lowest modes due to the weak attenuation.
As one scans temperatures within
the Fermi liquid regime, NMR peaks are expected to widen up to
a maximum width and then to become sharp again,
indicating the presence of a maximum value for the attenuation. 
The sole presence of a peak at higher temperatures will
provide evidence for this conservation in $2d$. The immediate
candidate systems for such experiments are helium films as
the ones pointed out here.\cite{Hallock,grey}
The existence of further consequences both in helium
layers and in $2d$ Fermi systems in general
is an open question.

The authors would like to thank useful discussions with A.J. Leggett.
P.F.F. is also grateful to B.I. Halperin and J.R. Engelbrecht,
to the hospitality and partial support of the Physics Department
at Boston College, and to FAPESP, the Research Foundation of the
State of S\~ ao Paulo, Brazil.

%

\end{document}